\documentclass[
    preprint,
    amsmath,
    amssymb,
    aps]{revtex4-2}
\usepackage{graphicx}

\usepackage{siunitx}
\usepackage{hyperref}

\DeclareSIUnit\angstrom{\text {Å}}
\usepackage{blindtext}

\begin{document}

    \title
    {
    Neural network analysis of neutron and X-ray reflectivity data: automated analysis using \emph{mlreflect}, experimental errors and feature engineering
    }
    
    \author{Alessandro Greco\textsuperscript{a}}
    \author{Vladimir Starostin\textsuperscript{a}}
    \author{Evelyn Edel\textsuperscript{a}}
    \author{Valentin Munteanu\textsuperscript{a}}
    \author{Nadine Rußegger\textsuperscript{a}}
    \author{Ingrid Dax\textsuperscript{a}}
    \author{Chen Shen\textsuperscript{b}}
    \author{Florian Bertram\textsuperscript{b}}
    \author{Alexander Hinderhofer\textsuperscript{a}}
    \author{Alexander Gerlach\textsuperscript{a}}
    \author{Frank Schreiber\textsuperscript{a}}
    \affiliation{
    [a] Institute of Applied Physics, Auf der Morgenstelle 10, University of Tübingen, 72076 Tübingen, Germany\\
    [b] Deutsches Elektronen-Synchrotron DESY, Notkestr. 85, 22607 Hamburg, Germany
    }

    \begin{abstract}

        This work demonstrates the Python package \emph{mlreflect} which implements an optimized pipeline for the automized analysis of reflectometry data using machine learning. The package combines several training and data treatment techniques discussed in previous publications. The predictions made by the neural network are accurate and robust enough to serve as good starting parameters for an optional subsequent least mean squares (LMS) fit of the data. It is shown that for a large dataset of 242 reflectivity curves of various thin films on silicon substrates, the pipeline reliably finds an LMS minimum very close to a fit produced by a human researcher with the application of physical knowledge and carefully chosen boundary conditions.
        
        Furthermore, the differences between simulated and experimental data and their implications for the training and performance of neural networks are discussed. The experimental test set is used to determine the optimal noise level during training. Furthermore, the extremely fast prediction times of the neural network are leveraged to compensate for systematic errors by sampling slight variations of the data.
         
    \end{abstract}

    \maketitle
    
    \section{Introduction}
        X-ray and neutron reflectometry (XRR and NR) are established surface scattering techniques that are routinely used to characterize solid and liquid thin films \cite{Tolan1999X, Holy1999High, Braslau1988Capillary,Russell1990X}.  They offer a non-invasive way to determine the structural, morphological or magnetic properties of a large variety of samples \cite{Neville2006Lipid, Skoda2017Simultaneous, Lehmkuehler2008Carbon} and can also be employed in real-time for in situ measurements \cite{Kowarik2006Real}. For decades, the conventional way to analyze reflectivity data has been the iterative least mean squares (LMS) or $\chi^2$ fitting of the data with a theoretical model \cite{Parratt1954Surface, Abeles1950La, Heavens1955Optical}. However, due to the well-known phase problem in scattering, the reconstruction of the scattering length density (SLD) profile from the reflectivity data is inherently ambiguous. This means that this method typically requires significant expertise and prior knowledge about the system, since for all but the simplest cases, there exist many possible solutions. Furthermore, even when the solution space is restricted, finding the global minimum is usually very time-consuming due to several local minima on the mean squared error (MSE) surface. For this purpose, various software packages have been developed over the years that use sophisticated minimization algorithms \cite{Bjoerck2007GenX, Kienzle2011Refl1D, Nelson2006Co, Nelson2019refnx, Danauskas2008Stochastic, Gerelli2016Aurore}. However, all of these approaches are iterative in nature and thus, usually computationally slow. Recently, machine learning-based methods have been proposed, that could avoid a lengthy search of the MSE surface, by providing an immediate guess for the thin film parameters that is already very close to the ground truth \cite{Greco2019Fast, Mironov2021Towards, Doucet2021Machine, Loaiza2021Towards, Greco2021Neural} or by encoding the reflectometry data into a latent space where the error surface does not have as many local minima \cite{Andrejevic2021Elucidating}.
        
        This paper demonstrates a Python-based reflectivity data analysis pipeline called \emph{mlreflect} that combines a fully-connected neural network regressor with several preprocessing and postprocessing steps to reliably predict the thickness, roughness and SLD of a thin film layer. While the principle of the neural network itself and the preprocessing have been discussed previously \cite{Greco2019Fast, Greco2021Neural}, here we focus on the differences between simulated and experimental data and show how this knowledge can be used to further optimize the obtained results. We tested the performance of the pipeline on a large experimental dataset of 242 XRR curves from different samples by comparing the result of the pipeline with manually supervised LMS fits that include physical knowledge and carefully chosen boundary conditions. This is a quantitative and qualitative difference compared to other similar studies, where most of the performance analysis is done with simulated data. In this context, we discuss the effect that experimental deviations from the theory can have on the training and prediction quality of the neural network. Using an example curve, we show how the extremely fast prediction speed of the neural network can also be leveraged to compensate for small experimental errors.
        
    \section{Description of the analysis pipeline}
                
        \begin{figure*}
            \centering
            \includegraphics[width=\linewidth]{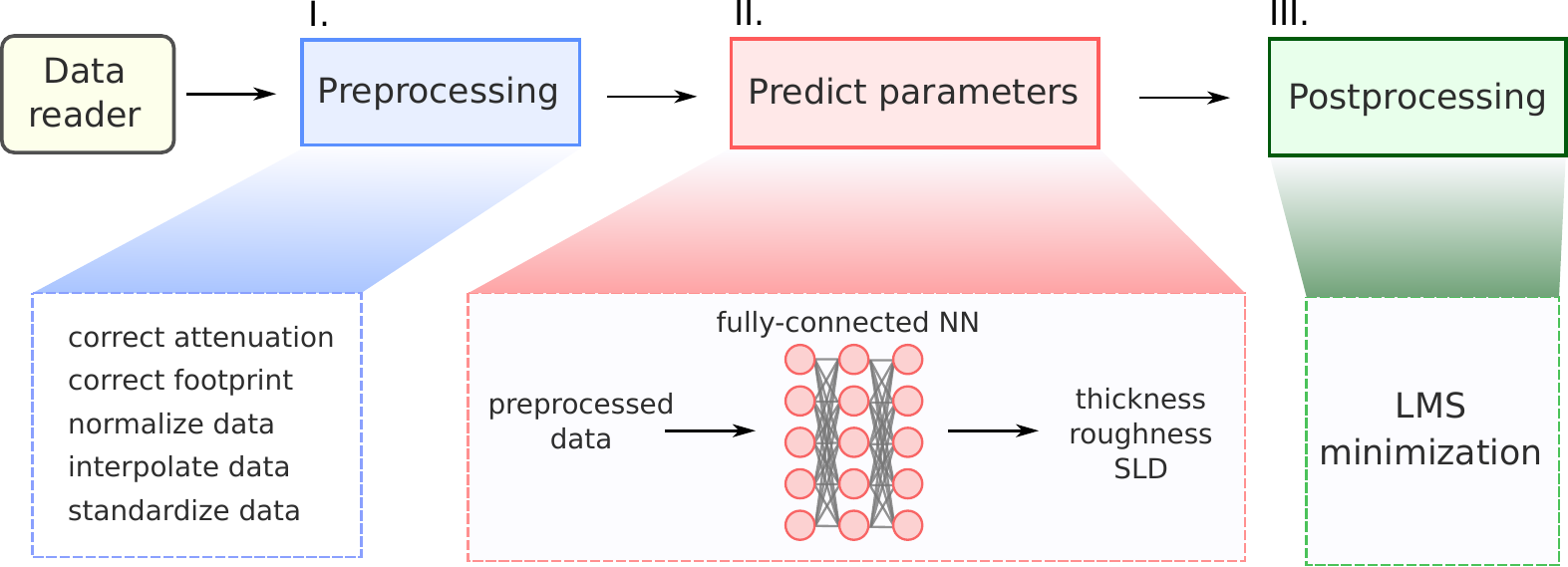}
            \caption{A schematic description of the analysis pipeline. The pipeline consists of three main steps: I. preprocessing, II. parameter prediction via the neural network and III. postprocessing. Step I. includes geometrical and other experiment-specific corrections. The data is also normalized, transformed into $q_z$ space, interpolated and standardized. In step II., the preprocessed data is fed into a trained, fully-connected neural network that yields an initial guess for the thin film parameters. During step III., this initial guess is used as starting parameters for a fast Levenberg-Marquardt fit that finds the nearest LMS minimum.}
            \label{fig:pipeline}
        \end{figure*}
    
        Our proposed analysis pipeline \emph{mlreflect} is fully written in Python and is available as open source on GitHub (\url{https://github.com/schreiber-lab/mlreflect}) and can also be downloaded directly from the Python Package Index (\url{https://pypi.org/project/mlreflect/}). The Supporting Information to this manuscript contains a step-by-step tutorial in the form of executable Jupyter notebooks (also available as PDF version). In addition, the tutorial, installation instructions and a full API documentation of the \emph{mlreflect} package are hosted on \url{https://mlreflect.readthedocs.io/en/latest/}. The neural network itself is implemented using TensorFlow \cite{Abadi2016TensorFlow}. It uses the matrix formalism implemented in the refl1d package \cite{Kienzle2011Refl1D} to simulate the reflectivity data. The workflow of the package can be conceptually separated into three steps: I. preprocessing, II. prediction and III. postprocessing, as depicted in \autoref{fig:pipeline}. Each of these steps is described in the following.
        
        During step I., the reflectivity data is automatically read from its raw format and several types of preprocessing procedures are applied. First, the raw data is converted into the standard $R(q_z)$ format where $R$ is the normalized reflected intensity and $q_z$ the momentum transfer vector along the surface normal. The preprocessing operations necessary depend on how the raw data is saved, but usually the data has to be corrected in some form. In our case, the raw data contains the reflected intensity at different scattering angles that must be corrected for the varying beam attenuation at different angles. Then, the intensity is corrected to account for the changing beam footprint on the sample at different angles, which amounts to a multiplication of the data with a geometric factor \cite{Gibaud1993correction}. Here we assume a flat sample and a beam with a Gaussian profile but, in principle, corrections for other sample or beam shapes can be implemented at this stage. The data is then normalized by dividing by the highest intensity value and transformed from angular space into $q_z$ space.
        
        After that, the intensity values are interpolated on a logarithmic scale to 109 equally-spaced $q_z$ points ranging from 0.02 to \SI{0.15}{\per\angstrom}, which corresponds to the input size of the neural network. Lastly, the data is standardized in the same way as was described in \cite{Greco2021Neural}, which ensures that each value of the input vector is on a similar scale. The effect on the general shape of the curves is comparable to multiplying the data with the inverse of the Fresnel reflectivity $R_\text{F}(q_z) \propto q_z^{-4}$, but, importantly, avoids the divergence for small values of $q_z$, i.e. close to and below the total reflection edge (TRE), where the kinematical approximation does not hold \cite{AlsNielsen2011Elements}.
       
        To obtain the initial parameter prediction (step II. in \autoref{fig:pipeline}) the preprocessed input vector is fed into the trained neural network model. The neural network is a fully-connected model that takes an input of 109 discrete intensity points and outputs 3 thin film parameters: the film thickness, the Névot-Croce film roughness \cite{Nevot1980Characterisation} and the real part of the scattering length density of the film. The model has 3 hidden layers with 512 neurons each. The training loss was calculated as the mean squared error between the normalized predicted and ground truth parameters. This architecture is similar to what has been described in the literature before \cite{Greco2019Fast, Greco2021Neural, Doucet2021Machine}, but to reduce training and inference times the number of parameters was reduced. The model was trained with 250000 simulated reflectivity curves with a batch size of 512. For every batch, uniform noise and curve scaling were applied to each curve to avoid overfitting as described before \cite{Greco2021Neural}. The optimal noise level during training was identified to be 0.3, which will be discussed in more detail later. Finally, the inputs were standardized as described above.
        
        The training data was generated assuming a sample structure consisting of a thin film on top of an oxide-capped silicon substrate with air as an ambient medium and with X-rays as the probe. The thin film parameters in the training data spanned a large range of 20--\SI{1000}{\angstrom} for the thickness, 0--\SI{100}{\angstrom} for the roughness and 1--\SI{14e-6}{\per\angstrom\squared} for the SLD. Furthermore, we restricted the roughness to values no higher than half the thickness since these scenarios are not well described by the theoretical model used. We note that a similar approach could easily be employed for neutrons or other sample structures  by retraining the neural network with different training data. We also expect this approach to work for a larger number of layers as long as the trained parameter space does not create too many ambiguous solutions, i.e., the number and range of fitting parameters should remain similar. For a larger parameter space, a larger $q_z$ range might be necessary to reduce ambiguity in the data. In our case, the $q_z$ range was limited to avoid conflicts with the Bragg peaks of organic molecules around \SI{0.3}{\per\angstrom} which are not described by the slab model.

        Lastly, during step III., the initially predicted thin film parameters are fed into an LMS minimizer to obtain the parameters which produce the best fit. Since the initial predictions are already very close to the ground truth, we chose a simple Levenberg-Marquardt minimizer \cite{More1977Levenberg} over a more powerful, but slower algorithm.

    \section{Performance test on thin films}
\begin{figure*}
            \centering
            \includegraphics[width=\textwidth]{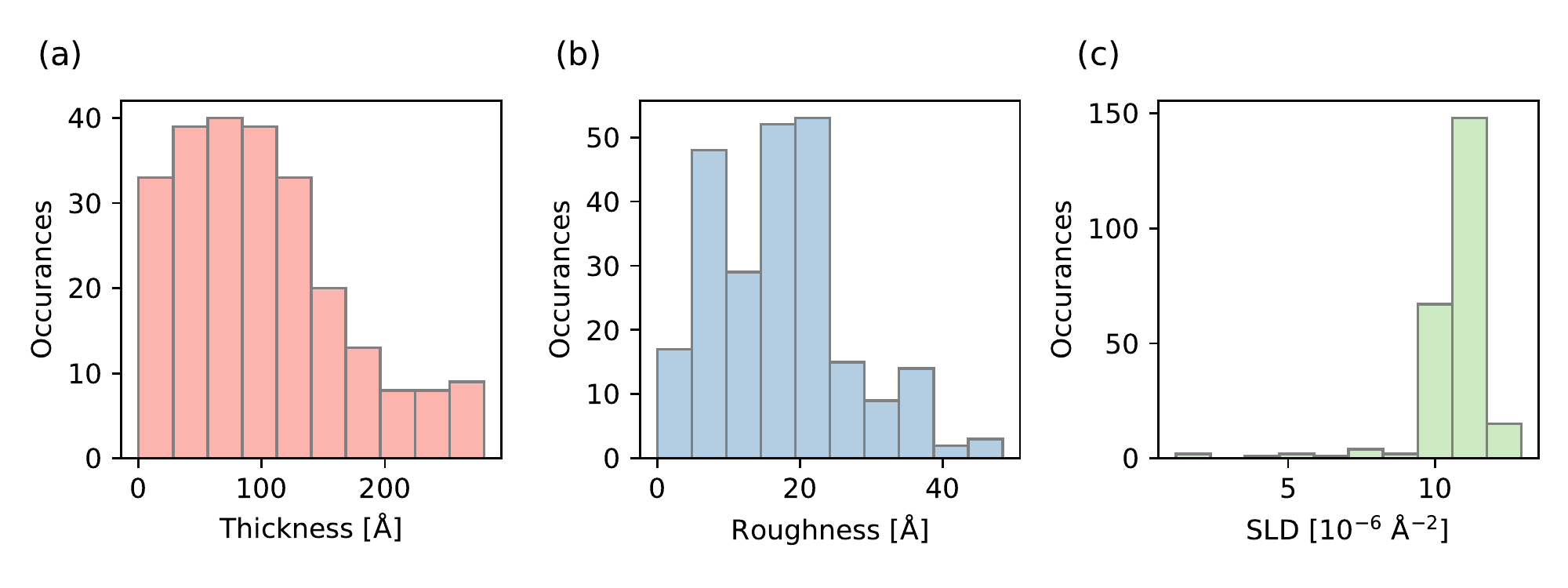}
            \caption{Ground truth distribution of the three sample parameters thickness (a), roughness (b) and SLD (c) within the experimental test set of 242 XRR curves. The parameters were obtained by a conventional LMS fit.}
            \label{fig:test_label_distribution}
        \end{figure*}

        The performance of the analysis pipeline was tested on 242 experimental XRR curves from in situ and ex situ experiments of 9 organic thin films on Si/SiO$_x$ (1--79 curves per sample at different thicknesses). The distributions of thickness, roughness and SLD of the film within this test set is shown in \autoref{fig:test_label_distribution}. The measurements were conducted using three different synchrotron radiation sources, i.e. the ESRF \cite{Smilgies2005Troika}, DESY \cite{Seeck2011high} and the SLS \cite{Patterson2005materials}, as well as using our own laboratory source. To obtain a benchmark, each reflectivity curve was first fitted on a logarithmic scale with an LMS fit based on the commonly used differential evolution algorithm \cite{Storn1997Differential} using manually chosen initial values and bounds for each parameter. The thin film model used for the fit was the same as what was used for training the neural network. In the following analysis, we assume that these manually fitted parameters represent the ``ground truth'' and thus the performance of our pipeline will be measured as the absolute error with respect to this ground truth.
        
        In the following, we compare the ground truth with prediction results of the neural network as well as the results of an automized subsequent LMS fit using the predicted parameters. Across all 242 curves, the neural network predictions have a median absolute error (median percentage error) of \SI{6.0}{\angstrom} (7.1\%) for the film thickness, \SI{2.0}{\angstrom} (12.4\%) for the interface roughness and \SI{0.72e-6}{\per\angstrom\squared} (6.8\%) for the SLD. This is a significant improvement to our first published model \cite{Greco2019Fast}, both on an absolute scale as well as on a relative scale, since the possible ranges for the thickness and roughness parameter have been greatly expanded. Thus, the network is generalized over a larger parameter space compared to previously published results. We note that since all of our data stems from organic thin films, the SLDs in the test set are mainly clustered around 10--\SI{13e-6}{\per\angstrom\squared}. Nevertheless, we assume that our results are not specific to the SLD range of the test data, since the network was trained equally with SLDs ranging from 1--\SI{14e-6}{\per\angstrom\squared}. We also want to highlight that the dataset also contains curves with a high roughness-to-thickness ratio where the Kiessig oscillations are strongly damped. Among the emerging solutions offered in this field, discussions about the performance on curves with little to no features are mostly absent. This is of course due to the challenge of extracting information from data that inherently contains less information. Yet, the network presented here also performs well on experimental data with high relative roughness.
        
        \begin{figure*}
            \centering
            \includegraphics[width=\linewidth]{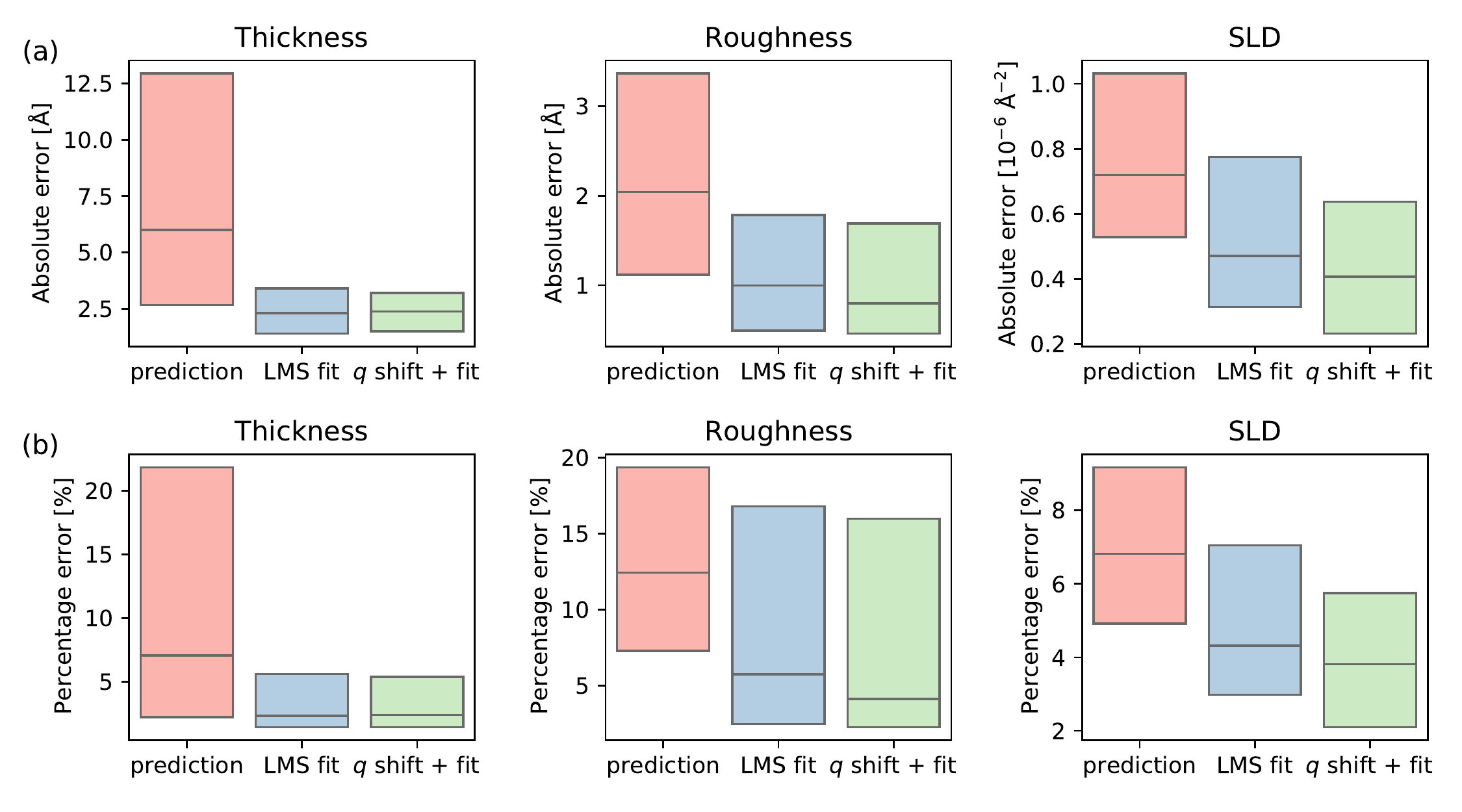}
            \caption{a) Box plot of the absolute errors for 242 measured reflectivity curves for each of the three predicted parameters.
            The upper and lower edge of the boxes represent the 1st and 3rd quartile with the horizontal line inside the boxes denoting the median. The blue boxes represent the error compared to the pure neural network predictions. The red boxes represent the error after applying a simple LMS minimization using the neural network predictions as starting parameters. The green boxes show the error for the case when a $q_z$ shift optimization has been performed before the LMS fit.
            b) The same box plots of the median error but as a percentage of the ground truth.\\
            All results were obtained for a training noise level of $n=0.3$.}
            \label{fig:performance}
        \end{figure*}
        
        The next step in the pipeline is to further refine these results via an LMS fit using the predictions from the neural network as starting parameters. Since the predictions are robust and already quite close to the ground truth there is no need for powerful but slow minimization algorithms such as genetic or differential evolution algorithms, which are normally employed to find the global minimum. Thus, finding the minimum takes only a fraction of a second per curve and can be fully automized. After this refinement procedure, the median absolute error (median percentage error) was even closer to the ground truth with \SI{2.3}{\angstrom} (2.3\%) for the thickness, \SI{1.0}{\angstrom} (5.8\%) for the roughness and \SI{0.47e-6}{\per\angstrom\squared} (4.3\%) for the SLD. A comparison of the error distributions before and after refinement is shown in \autoref{fig:performance}. A detailed breakdown of the prediction error with respect to each parameter can be found in Figures S2--S10 of the Supporting Information.
        
        The residual error can be attributed to the fact that every fit has a finite accuracy and hence the ground truth itself contains a certain error. We roughly estimate this error to be at least $\pm 10\,\%$ for each parameter, which would be comparable to the reported error of the neural network. Thus, these results show that the analysis pipeline as described above performs similarly to a human researcher in most circumstances. Furthermore, it is important to note that the results were obtained much faster than via a human-guided fit. Excluding the time it took to train the neural network (about 20 minutes for a given sample structure), the initial parameter predictions of the 242 curves were obtained after only 1 second with about 2 additional minutes for the further refinement steps, resulting in a total fitting time of about 0.4s per curve. In contrast, producing the ground truth fits took about 6 hours because of the need to carefully select fitting boundaries to prevent the fit from running into non-physical minima.

    \section{Differences between simulated and experimental data}
        A well-known property of artificial neural networks is that they require large amounts of representative training data to learn a generalized model and not overfit the training set. In the context of the work presented here, i.e. supervised learning using scattering data, this would mean acquiring thousands of scattering patterns from a large variety of different samples and analyzing them manually to create the training set. Since this is a quite time-consuming and challenging task, neural network models in the field of scattering physics are typically trained with simulated data based on well-established theoretical models. In most cases, the simulation is additionally modified with certain artifacts, such as noise, to better mimic experimental conditions. However, to what degree this is necessary is difficult to estimate since the only available metric is typically the performance on other simulated data (validation loss), which is expected to decrease with increasing perturbations.
        
        \begin{figure*}
            \centering
            \includegraphics[width=3.5in]{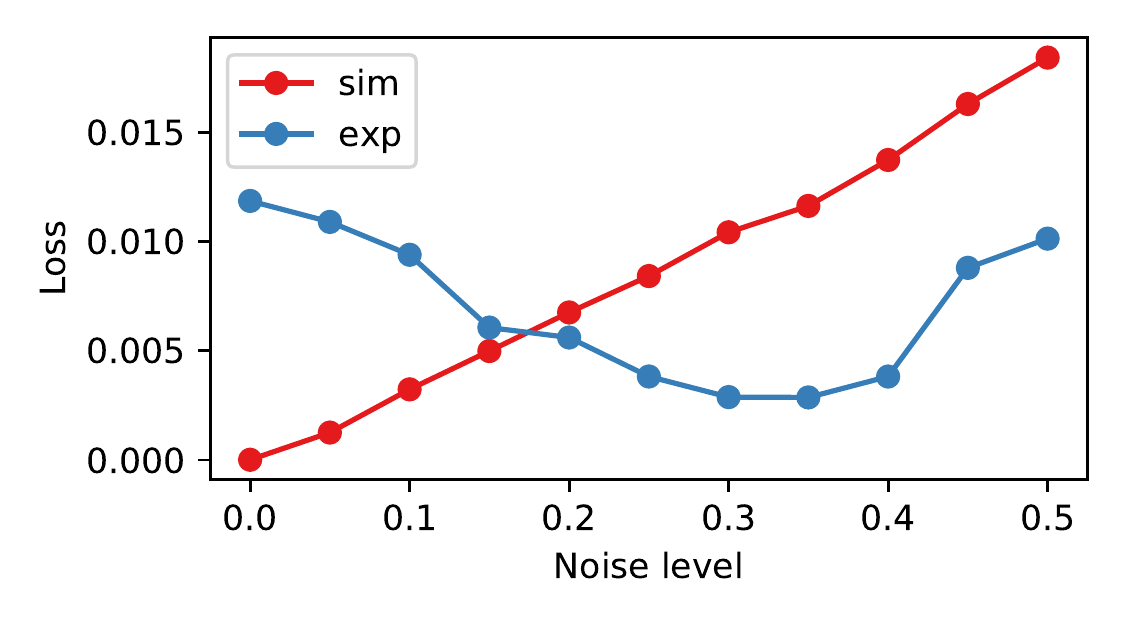}
            \caption{Comparison of the loss calculated from a simulated test set (100000 curves) and an experimental test set (242 curves) for different levels of uniform noise that were applied to the training data. For each noise level a separate model was trained. With increasing noise level, the loss from the simulated data increases linearly while the loss from the experimental data shows a clear minimum at noise levels 0.3--0.35.}
            \label{fig:noise_levels}
        \end{figure*}
        
        In this study, we investigated how applying uniform noise to the training data affects the neural network performance on our large experimental dataset of 242 curves. We trained 11 copies of the same neural network (as described above) with training data with different noise levels $n$ where each data point $R_i^*$ in the noisy curve was sampled uniformly between the values $R_i(1-n)$ and $R_i(1+n)$. Thus, $n$ denotes the maximum relative change of a given data point $R_i$ of a given simulated curve. The $n$ for each trained model ranged from 0 to 0.5 in 0.05 increments. It is important to note that the applied uniform noise is not meant to model a specific physical noise type, such as Poisson noise for counting statistics. Rather, uniform noise was chosen as a $q$-independent catch-all noise that affects the whole curve equally and thus, makes the neural network robust against errors across the entire $q$ range.
        
        \autoref{fig:noise_levels} shows a comparison of the losses calculated with a simulated test set as well as with the experimental test set for each model. Since the loss is calculated as the mean squared error of all three (normalized) sample parameters, it is a unitless measure for the accuracy of the model. For $n=0$, the simulated test set shows a loss close to zero ($\sim 10^{-7}$), whereas the loss based on the experimental data is about 5 orders of magnitude higher. This shows that without any noise, the neural network significantly overfits the simulation and thus performs suboptimally on real data.  As expected, the loss of the simulated data increases monotonically with increasing noise. However, the performance on the real data improves significantly with increasing noise up to a noise level of 0.3--0.35. Beyond this, even higher noise levels seem to again worsen the performance. This very clearly demonstrates that there exists an optimal noise level for which the added noise acts as an effective regularization technique that prevents overfitting. If the noise level is too high, however, the consequent lack of information is likely detrimental to the training. Thus, we identified $n=0.3$ to be the ideal noise level for data similar to our testing set, which notably contains data from different X-ray sources. Furthermore, Figure S1 of the Supporting Information shows that the optimal training noise does not significantly change for subsets with different noise levels (0.1--0.5) within the experimental test set. Thus, we set the default value of the noise level in our analysis pipeline to 0.3.  Datasets that differs significantly from our test set in terms of experimental artifacts might of course produce slightly different results, although we expect the general trend to be the same. This highlights the importance of having a large experimental test set with representative experimental artifacts since metrics only based on simulated data are clearly not sufficient to evaluate the training progress.

    \section{Influence of systematic measurement errors}
        All reflectometry measurements are performed with a finite accuracy due to various error sources. These errors are detrimental to the experiment and can impede the extraction of information from the data and therefore should be avoided or minimized as much as possible. However, a finite error inevitably remains for every measurement. Among the possible statistical errors are the signal-to-noise ratio, the angular resolution of the diffractometer and the spectral resolution of the source. Among the systematic errors are, for example, the convolution of the data with the slit functions, the accuracy of the sample alignment and the accuracy of the footprint correction (i.e. how accurate the beam and sample shape can practically be determined).
        
\begin{figure*}
            \centering
            \includegraphics[width=3.5in]{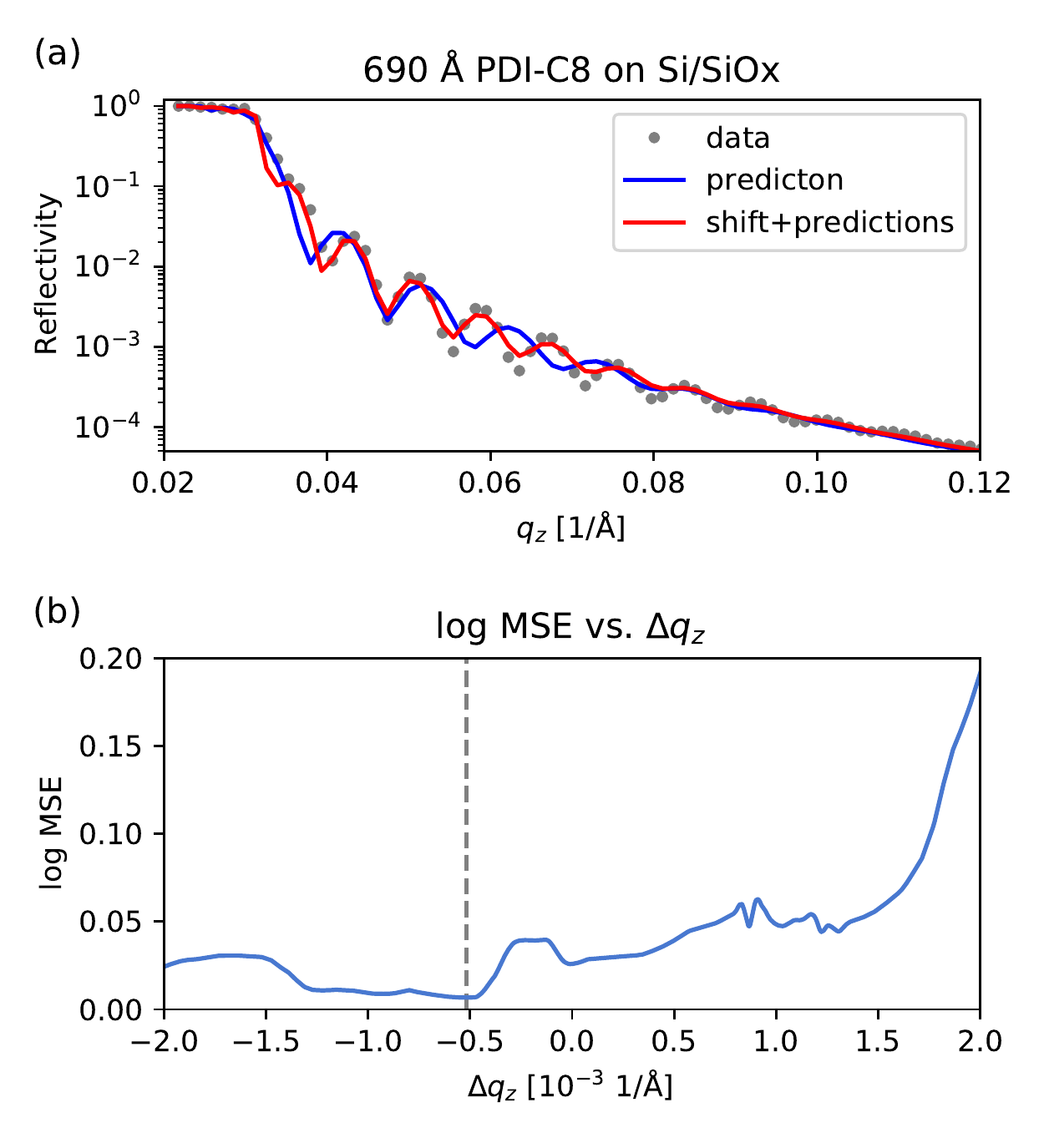}
            \caption{a) Comparison of the neural network predictions from reflectivity data from a \SI{690}{\angstrom} thick PDI-C8 film on Si/SiO$_x$. The blue curve shows the native prediction whereas the red curve shows the prediction after the data was shifted by $\Delta q_\text{min} = \SI{5.2e-4}{\per\angstrom}$ before the interpolation step. It is apparent that the latter is in much better agreement with the data. b) Shows the log MSE between the predicted curve and the data for different $\Delta q_z$. The minimum MSE at $\Delta q_\text{min}$ is indicated by the dashed line.}
            \label{fig:shift_comparison}
        \end{figure*}

        Having imperfect data obviously impacts the analysis, since the data deviates from the ideal physical model it is compared with. Since the neural network model presented here is trained to solve a very particular task that assumes well-defined data, these errors can negatively impact the prediction quality. In general, it is easier to make the neural network robust against statistical errors by introducing them during training, as described before. However, sometimes systematic errors, such as a small misalignment can also seriously misguide the ML prediction, as shown in \autoref{fig:shift_comparison}. Therefore, it would be useful to correct or compensate some of these errors during inference time after the data has been acquired.
        
        As a solution, we propose an automated method for sampling through slight variations of the data, exploiting both the sensitivity and speed of our neural network model. Since the neural network assumes data that conforms to an idealized physical model, it might fail if the data contains anomalies with respect to that model. Since predictions with the neural network are very fast, it is possible to scan through thousands of modified reflectivity curves within less than a second. For each of these variants, the log MSE between the data and the predicted curve can be calculated and only the one with the lowest error is subsequently selected. We demonstrate an implementation of this method that identifies small systematic alignment errors and automatically applies an appropriate shift to the data.
        
        \autoref{fig:shift_comparison}a shows an XRR measurement of a \SI{690}{\angstrom} thick N,N'-Dioctyl-3,4,9,10-perylene\-di\-carboximide (PDI-C8) film on Si/SiO$_x$ which was measured and tested in addition to the 242 test curves. Here, in contrast to the previously shown test set, the normal pipeline as described above did not converge to the correct minimum. The reason for this is the much higher thickness of the film, which leads to denser Kiessig oscillations in the data. This, in turn, creates many narrow minima on the MSE surface for the LMS algorithm to get trapped in. As a result, the neural network prediction needs to be even closer to the ground truth for the subsequent fit to converge. \autoref{tab:shift_comparison} shows the predicted thin film parameters in comparison to the ground truth. A possible reason for the suboptimal neural network prediction might be small imperfections in the data due to finite measurement errors, such as a small variation in sample alignment. In regions of high derivatives, even a small shift of the data along the $q_z$ axis can lead to strong differences in the observed intensities at a given $q_z$ value, even on a logarithmic scale. Of course, if the data has dense oscillations, this effect becomes more pronounced. For models trained on simulated data, this can be critical, since normally a substantial change of certain input neurons, especially near the TRE, corresponds to important information and will be interpreted by the network accordingly. To check whether this can be remedied, we shifted the $q_z$ values during the interpolation step by a small value $\Delta q_z$ and repeated the prediction. This was done 1000 times with randomly sampled $\Delta q_z$ ranging from \SI{-1e-3}{} to \SI{1e-3}{\per\angstrom}. Then, for each prediction, the quality of the prediction was evaluated by calculating the log MSE between the corresponding simulation and the measured curve.
        
        %since the initial guess of the neural network was not close enough
        
        \begin{table*}
            \centering
            \caption{Predicted and fitted thin film parameters based on the reflectivity data of a PDI-C8 film on Si/SiO$_x$ (shown in \autoref{fig:shift_comparison}). The ground truth labels were obtained via a manually supervised LMS fit. After applying the described $q_z$ variation, the prediction results improved significantly. A subsequent LMS refinement only led to comparatively small improvements.}
            \begin{tabular}{lrrr}
                \hline
                {} &  Thickness [Å] &  Roughness [Å] &  SLD [$10^{-6}$\si{\per\angstrom\squared}] \\
                \hline
                ground truth             &          688.3 &           27.1 &                      10.5 \\
                prediction               &          536.7 &           30.3 &                      11.2 \\
                shift + prediction       &          690.8 &           31.0 &                      11.0 \\
                shift + prediction + fit &          690.5 &           27.5 &                      10.8 \\
            \end{tabular}
            \label{tab:shift_comparison}
        \end{table*}
        
        When plotting the log MSE between the prediction and the input against $\Delta q_z$ (\autoref{fig:shift_comparison}), we observed a value $\Delta q_\text{min} = \SI{5.2e-4}{\per\angstrom}$ for which the log MSE shows a clear minimum. From \autoref{fig:shift_comparison}a it is apparent that the predicted curve based on the shifted data shows much better agreement with the data than the normal prediction. The corresponding predicted parameters for $\Delta q_\text{min}$ are shown in \autoref{tab:shift_comparison} and are much closer to the ground truth values (comparable to values shown in the previous section). This indicates that there exists a certain shift $\Delta q_\text{min}$ that can (at least partially) compensate for the experimental error. This is especially valuable since it allows the pipeline to continue with the LMS refinement step, which ultimately leads to a near-perfect fit.
        
        It is interesting to note that $\Delta q_\text{min}$ is very small, corresponding to a change of the angle of incidence of only about \SI{4e-3}{} degrees for a wavelength of \SI{1.54}{\angstrom}. It seems intuitive that such a small shift in the data could be caused by a variety of the above mentioned error sources. However, although $\Delta q_\text{min}$ is seemingly small, due to the high derivatives close to the TRE and the Kiessig fringes, shifting the data by $\Delta q_\text{min}$ still has a noticeable effect on each data point. For conventional LMS fitting, this might not seem critical at first, since the MSE surface likely has a minimum close to the real one in terms of the film thickness. However, for the roughness and density parameters this might not be the case and thus, most fitting programs allow the user to manually shift the data if necessary.
        
        While in principle any type of modification like this could be conceivably applied to the data to scan for the lowest MSE, we observed significantly better results with this method rather than, for example, adding Gaussian noise. This is because a translation of the curve preserves most of the information in the data while still varying every data point, in contrast to Gaussian sampling which is $q$-independent and inevitably destroys information.
        
        To test the stability of this method, we applied the $\Delta q_z$ sampling procedure to all 242 curves discussed in the previous section (where the pipeline already succeed) and compared the results with the original mean absolute error. When looking at \autoref{fig:performance}, it becomes clear that scanning for $\Delta q_\text{min}$ did not harm the mean absolute error, but instead even improved the results slightly for all three parameters. While the log MSE of the predictions is already very close to the minimum, most of the data likely still has a finite alignment error which, however, was not sufficient to affect the prediction. Hence, this could still be compensated by applying a small shift, ultimately leading to an even better fit. Because this screening for $\Delta q_z$ yielded significant improvements on some data and was relatively fast, we decided to routinely add this to the analysis pipeline.

    \section{Fourier transforms as a method for feature engineering}
        The specular reflectivity from a single layer on a substrate well above the critical angle can be approximately described by
        
        \begin{equation}
            R(q_z) = R_\text{F}(q_z) \left| \int_{-\infty}^\infty \frac{d\rho (z)}{dz} e^{iq_zz} dz\right|^2
        \end{equation}
        
        \noindent
        i.e. the product of the Fresnel reflectivity from a flat surface and the squared Fourier transform of the SLD contrast of the sample along the surface normal \cite{AlsNielsen2011Elements, Sivia2011Elementary}. Although the phase of the Fourier transform is lost by taking its absolute square, the inverse Fourier transform of $R(q_z)/R_\text{F}(q_z)$ still carries some important information such as the frequency of the Kiessig oscillations (and thus the film thickness). As a result, performing an inverse Fourier transform on the reflectivity data presents itself as an obvious way to create additional input features that may facilitate the neural network training.
        
        To test this hypothesis, we trained a neural network model with an additional preprocessing step before the first layer that performs a Fast Fourier transform on the standardized input and adds the real and imaginary Fourier components to it, leading to a input layer size of 219 neurons. All other model parameters and training ranges were kept the same as described above. When testing the trained model on the 242 experimental curves, we found that the model performed similarly to the model without the added Fourier transform. The median absolute error (median percentage error) was \SI{6.2}{\angstrom} (8.9\%) for the film thickness, \SI{2.3}{\angstrom} (13.3\%) for the interface roughness and \SI{0.76e-6}{\per\angstrom\squared} (7.2\%) for the SLD, which is 4\%, 19\% and 6\% higher, respectively, compared to the base model.
        
        From this we conclude that the base model (without the added Fourier transform) had likely already learned to implicitly extract all available frequency information from the data and adding the Fourier components explicitly does not lead to a better training result. Furthermore, the reason why the results are slightly worse when the Fourier transform is added might be attributed to the increased number of trainable parameters due to the larger number of neurons in the model. Thus, more parameters need to be optimized to achieve the best training result, which is generally a more difficult task. For these reasons and the added computational requirements during both training and inference time, we decided not to include the Fourier transform layer into the default neural network layer of our analysis pipeline. Nevertheless, we do not rule out that a suitable implementation of the Fourier transform could be beneficial for certain scattering geometries.

    \section{Conclusion}
        We demonstrated an optimized analysis pipeline, \emph{mlreflect}, based on machine learning for the automated analysis of reflectivity data. We tested our pipeline on a large dataset of 242 XRR curves, containing in situ and ex situ measurements of organic thin films on Si/SiO$_x$ substrates, where it showed a performance comparable to a manually supervised least mean squares fit for most of the data. Therefore we conclude that \emph{mlreflect} is a useful tool for the automated pre-screening or even on-the-fly analysis of reflectivity data.
        
        We also discussed that for the effective evaluation of trained machine learning mo\-dels a sufficiently large experimental dataset is necessary. Most studies so far mainly focus on the performance of the model with regards to simulated data and include only few, if any, experimental test data. However, this may be misleading, since our results clearly show that the performance on simulated data cannot easily be generalized to experimental conditions.
        
        Furthermore, we showed the influence of possible systematic errors (such as misalignment) on the reflectivity data and how the prediction speed of the neural network model can be exploited to improve the overall performance by transforming the data slightly. Our results highlight the necessity of accounting for these differences between simulated theoretical models and real data in order to obtain stable results.
        
        Although the results shown here are demonstrated with systems of one layer on a Si/SiO$_x$ substrate, the demonstrated neural network model could easily be retrained to determine any single layer of any sample structure. While determining multiple layers at once is in principle possible and has been demonstrated before, this type of neural network architecture might not be ideal to tackle this type of inverse problem with multiple solutions since they map exactly one solution to a given input. Therefore, architectures that yield probabilities as an output might be more suitable for multi-layer problems.

    \section{Acknowledgments}
        This research was funded by the German Federal Ministry for Science and Education (BMBF).
        
        We acknowledge DESY (Hamburg, Germany), a member of the Helmholtz Association HGF, for the provision of experimental facilities. Parts of this research were carried out at PETRA III and we would like to thank Rene Kirchhof for assistance in using photon beamline P08. Beamtime was allocated for proposal II-20190761. This research was also supported in part through the Maxwell computational resources operated at DESY with the assistance of André Rothkirch and Frank Schlünzen.
        
        We acknowledge the Paul Scherrer Institut, Villigen, Switzerland for provision of synchrotron radiation beamtime at the Materials Science beamline of the SLS.
        
        We acknowledge the European Synchrotron Radiation Facility for provision of synchrotron radiation facilities at beamline ID10b.
        
        We thank the Deutsche Forschungsgemeinschaft (DFG) for financial support.
        
        Supported by the German Research Foundation through the Cluster of Excellence “Machine Learning – New Perspectives for Science”.
    
    \bibliography{main}
    
\pagebreak
    
\part*{Supporting Information}

    \renewcommand{\thefigure}{S\arabic{figure}}
    \setcounter{section}{0}
    \setcounter{figure}{0}

    \section{Optimal training noise levels}
        
        Figure 4 of the main manuscript shows the loss on the experimental dataset of 242 curves for 11 different neural network models where the training data was modified with different amounts of uniform noise. The results show that there seems to be an optimal noise value of about 0.3 where the loss for the experimental data has a minimum.
        
\begin{figure*}
            \centering
            \includegraphics[width=\textwidth]{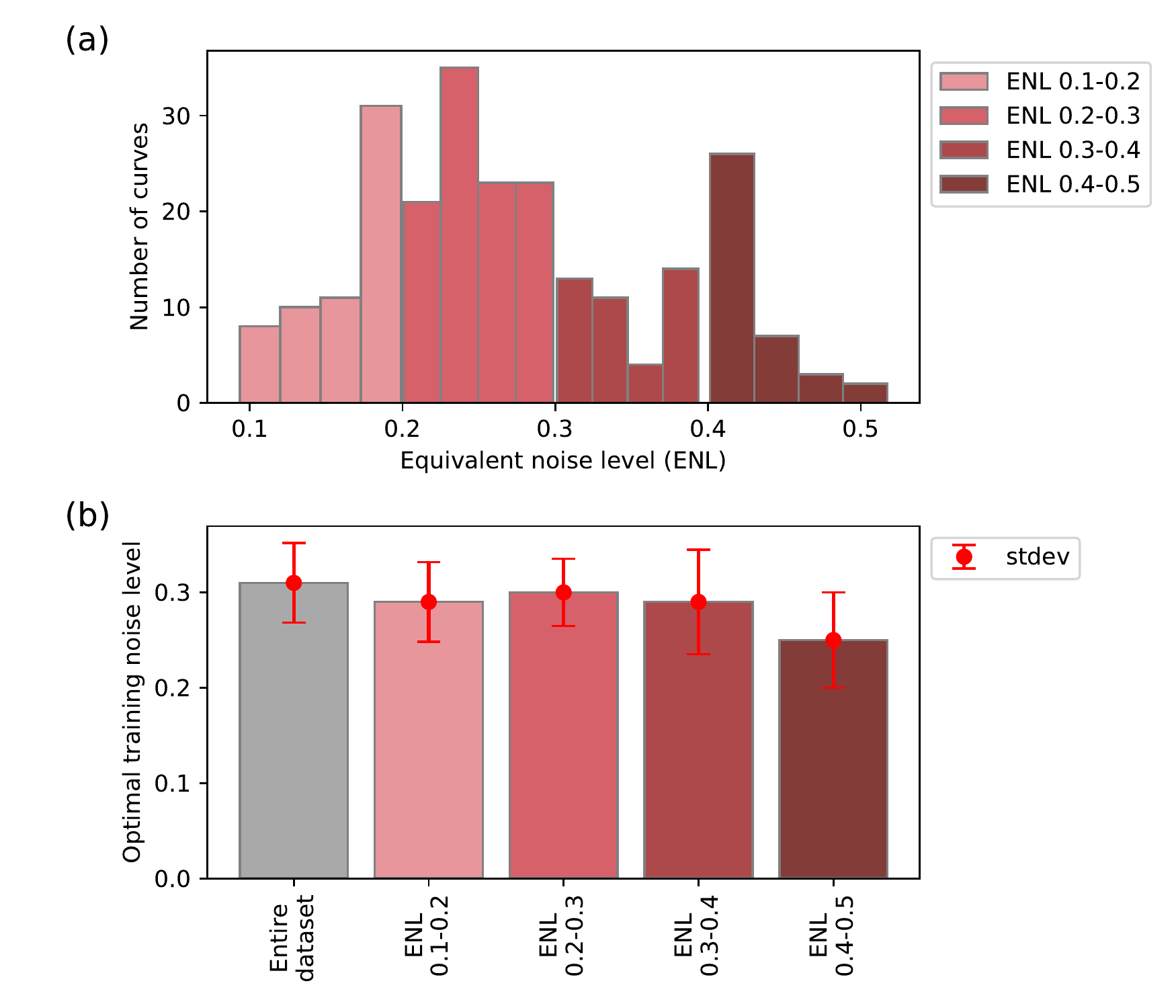}
            \caption{(a) Distribution of the equivalent noise level (ENL) in the experimental testing dataset of 242 XRR curves. The dataset was split into four categories with varying ENLs to test each separately. (b) Optimal training noise for different ENLs in the training data. The optimal level for the entire dataset corresponds to the minimum shown in Figure 4 of the main manuscript. The error bars represent the standard deviation of five independent training repetitions.}
            \label{fig:noise_levels2}
        \end{figure*}

        An interesting question arises about how this value is related to the amount of noise in the experimental data. To investigate this, the test data was separated into four groups with varying amounts of noise. While the noise in the data is not uniformly distributed, an equivalent noise level (ENL) can be calculated by subtracting the ground truth fit from the data and taking the absolute mean of all data points. \autoref{fig:noise_levels2}a shows the distribution of the ENL across the entire dataset and how the distribution was split into the four subsets with a different ENL. \autoref{fig:noise_levels2}b shows the optimal training noise (for which the loss had a minimum) for each of the four categories as well as the entire dataset. The error bars represent the standard deviation of five independent training repetitions. Evidently, the ENL of the data does not seem to have a strong influence on the optimal noise level except for the 0.4--0.5 category, where it is slightly lower. This is due to the main source of error in the data not being statistical (e.g. Poisson noise), but rather systematic in nature (e.g. the fit does not fully describe the data). Since the role of the uniform noise on the training data is not to mimic the noise in the data, but to account for these systematic deviations, the entire dataset benefits from a similar training noise level.
        
        However, for data with significantly higher statistical noise than our dataset, it could be possible that optimal training noise is different.
    
    \section{Detailed prediction error histograms}
\begin{figure*}
            \centering
            \includegraphics[]{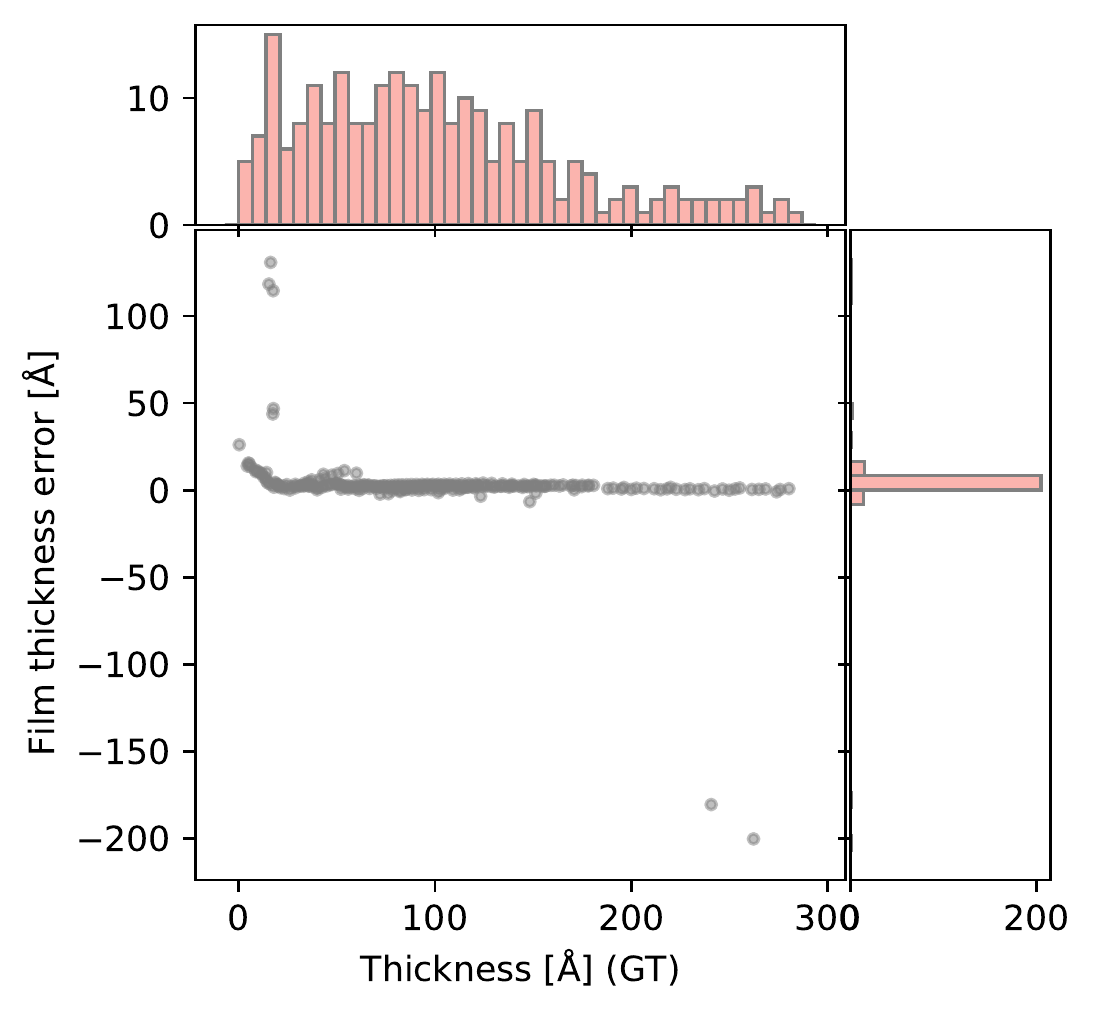}
            \caption{Distribution of the absolute thickness error from the full pipeline fit with respect to the ground truth (GT) thickness. Each dot represents a single curve in the testing dataset.}
            \label{fig:thickness_error_hist_thickness}
        \end{figure*}

\begin{figure*}
            \centering
            \includegraphics[]{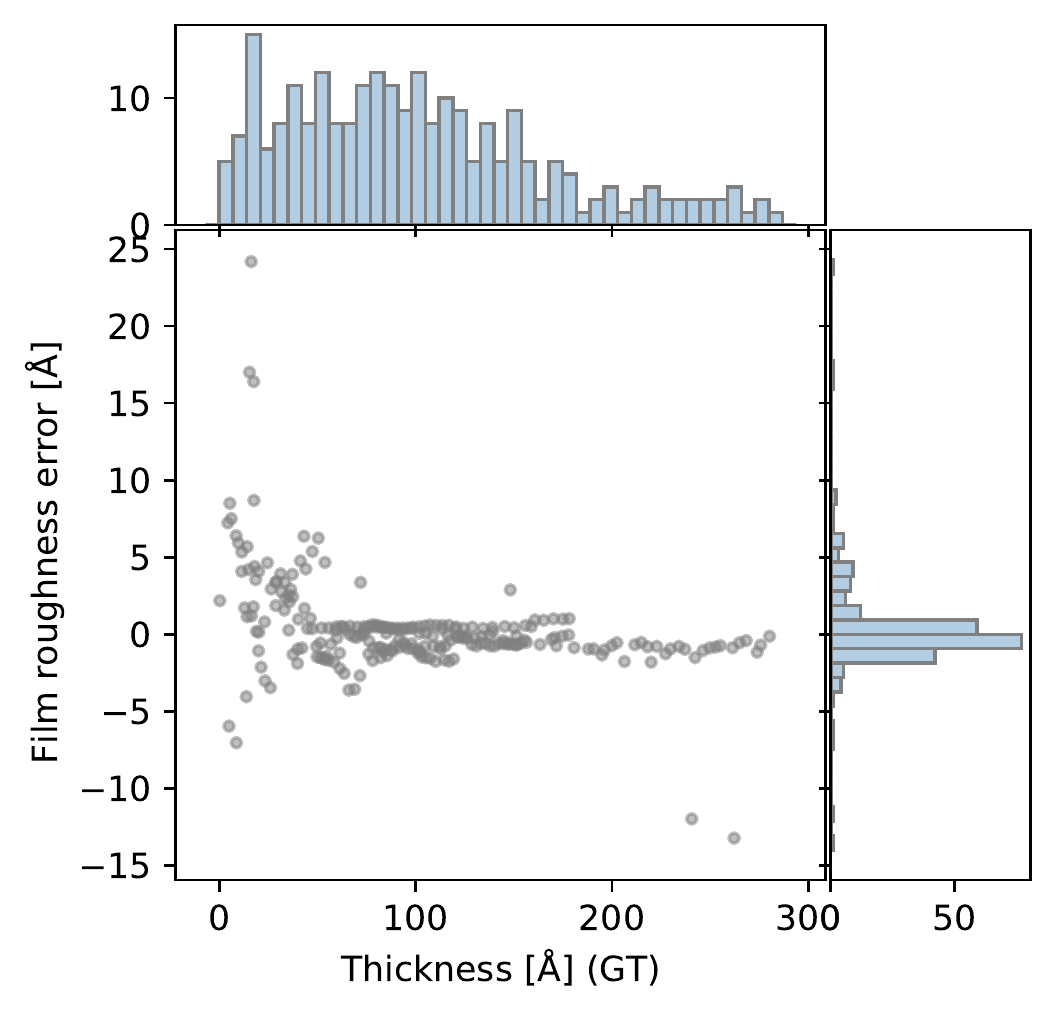}
            \caption{Distribution of the absolute roughness error from the full pipeline fit with respect to the ground truth (GT) thickness. Each dot represents a single curve in the testing dataset.}
            \label{fig:thickness_error_hist_roughness}
        \end{figure*}

\begin{figure*}
            \centering
            \includegraphics[]{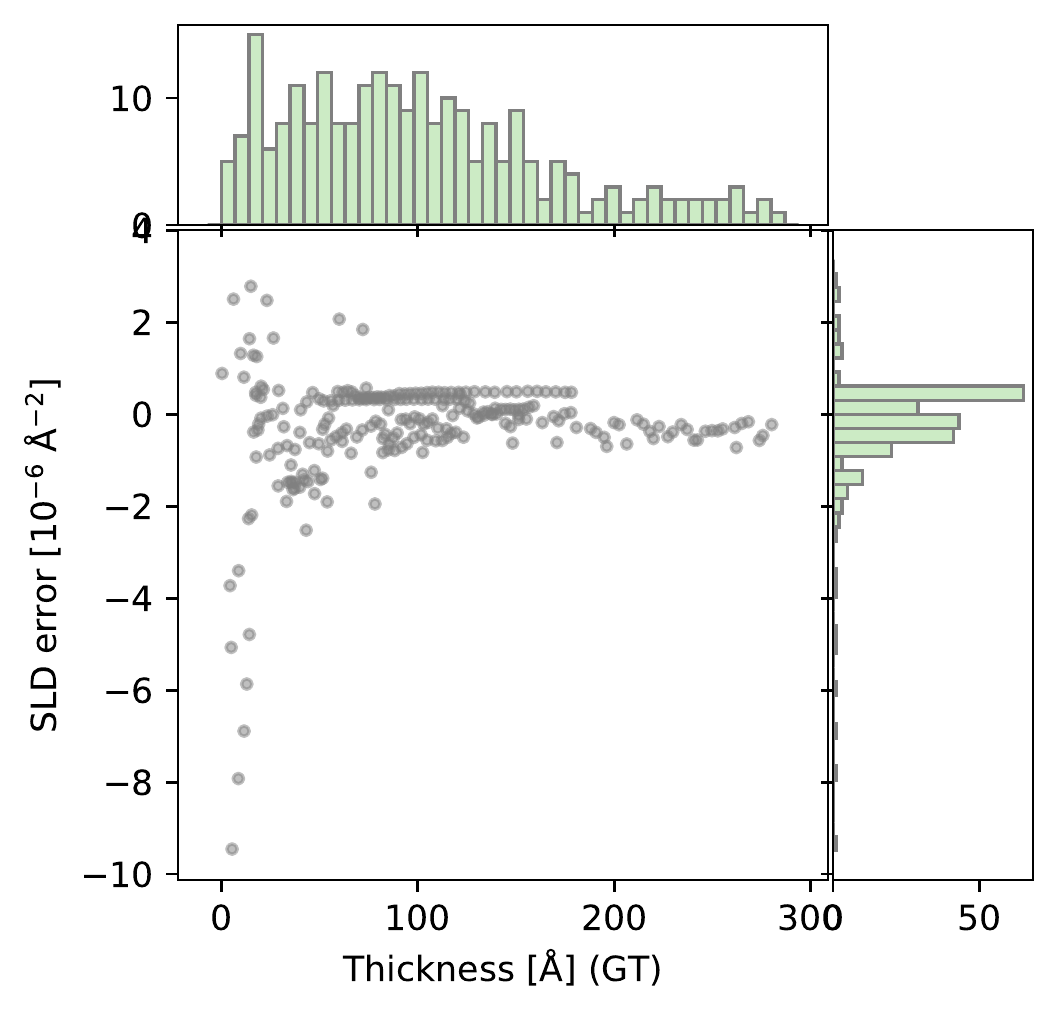}
            \caption{Distribution of the absolute SLD error from the full pipeline fit with respect to the ground truth (GT) thickness. Each dot represents a single curve in the testing dataset.}
            \label{fig:thickness_error_hist_sld}
        \end{figure*}

\begin{figure*}
            \centering
            \includegraphics[]{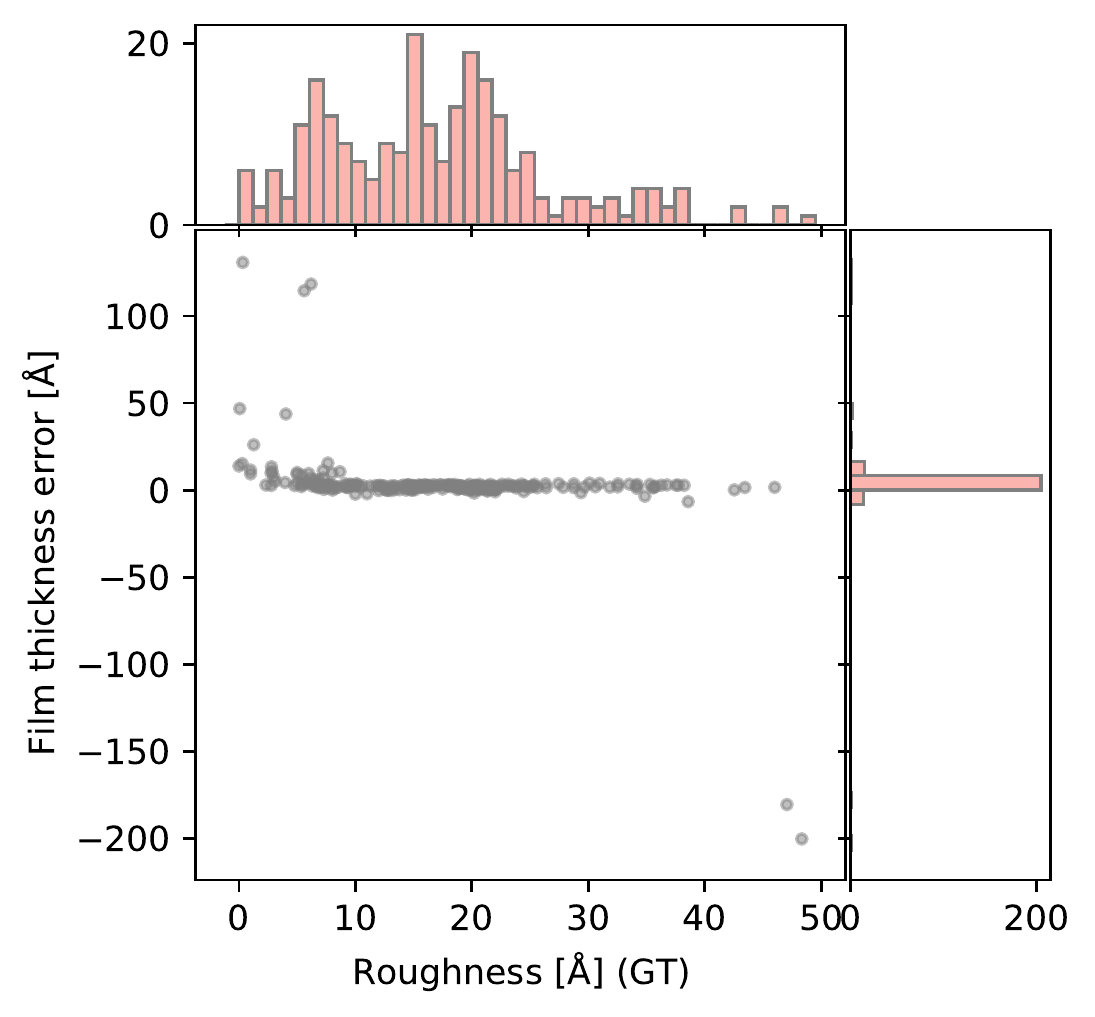}
            \caption{Distribution of the absolute thickness error from the full pipeline fit with respect to the ground truth (GT) roughness. Each dot represents a single curve in the testing dataset.}
            \label{fig:roughness_error_hist_thickness}
        \end{figure*}

\begin{figure*}
            \centering
            \includegraphics[]{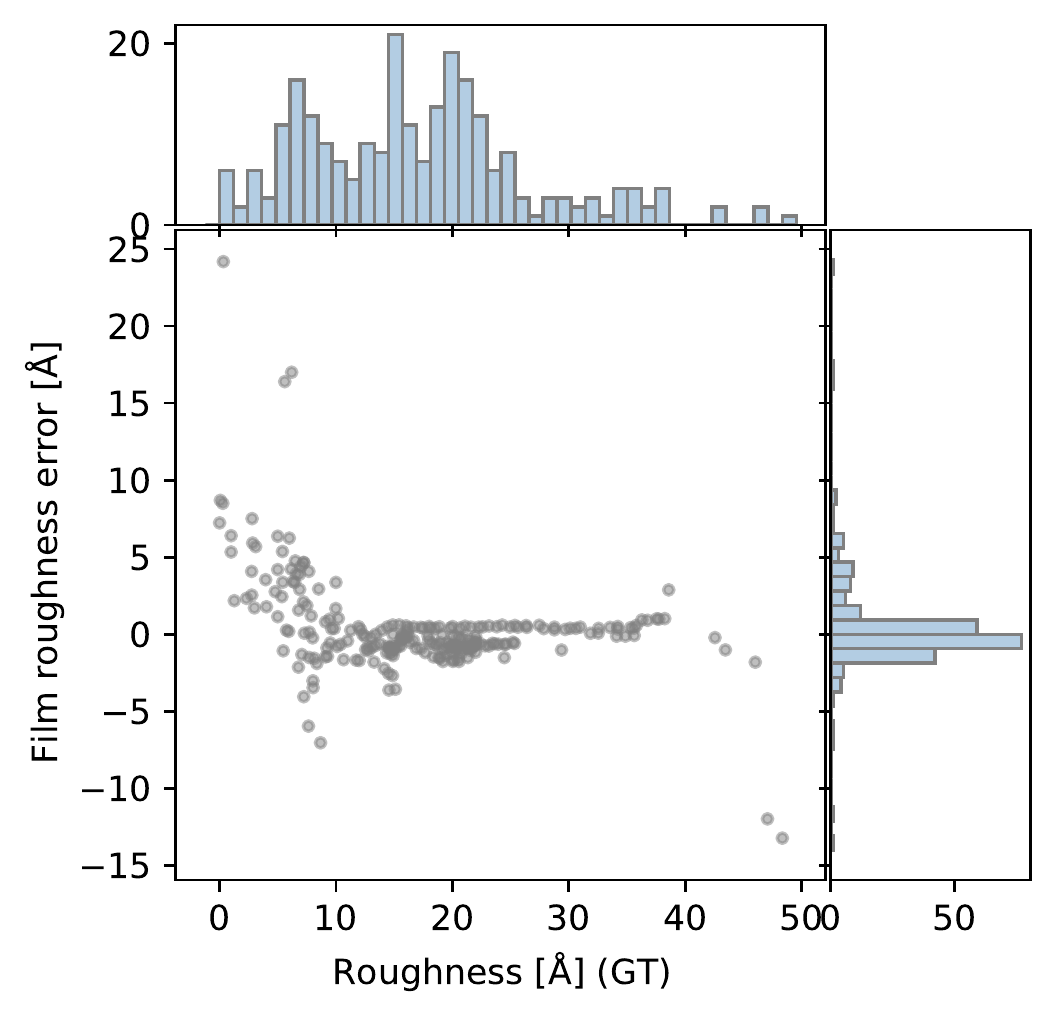}
            \caption{Distribution of the absolute roughness error from the full pipeline fit with respect to the ground truth (GT) roughness. Each dot represents a single curve in the testing dataset.}
            \label{fig:roughness_error_hist_roughness}
        \end{figure*}

\begin{figure*}
            \centering
            \includegraphics[]{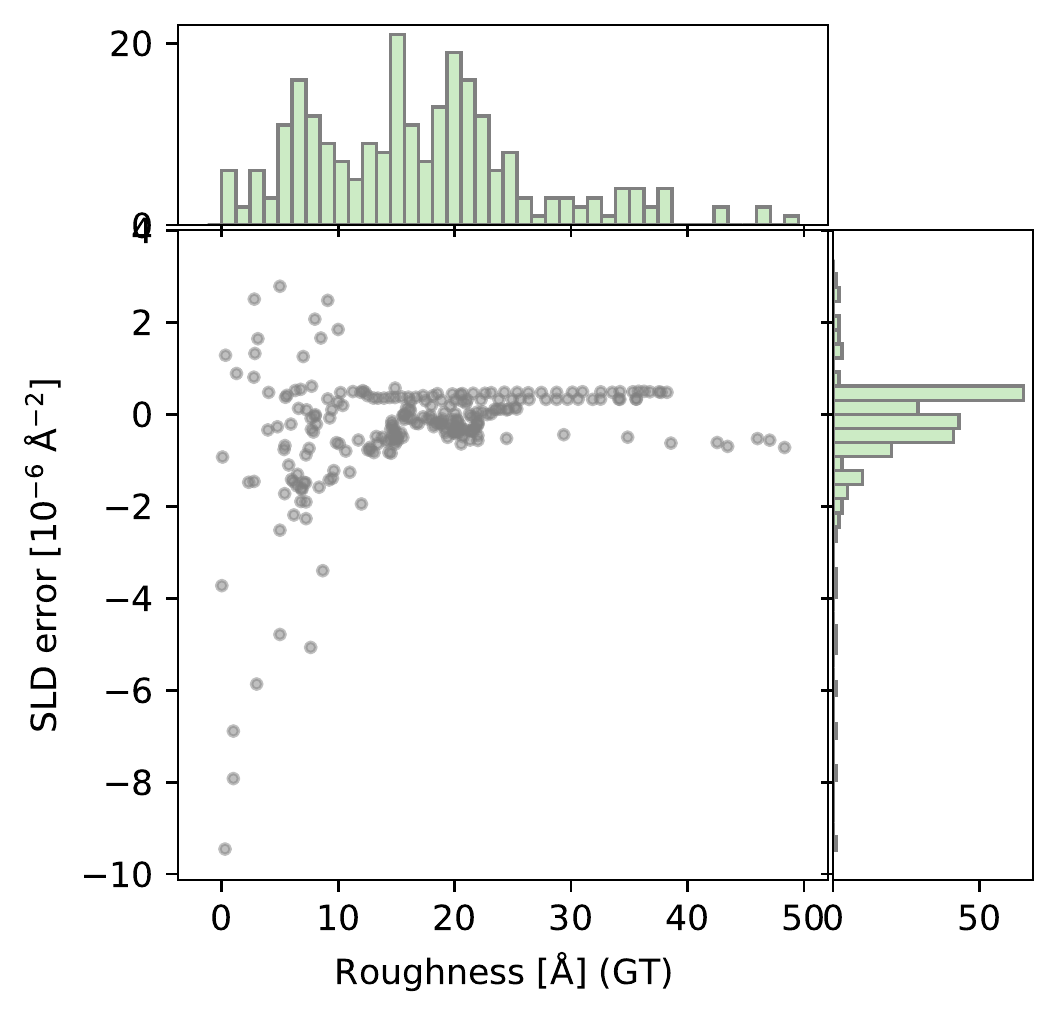}
            \caption{Distribution of the absolute SLD error from the full pipeline fit with respect to the ground truth (GT) roughness. Each dot represents a single curve in the testing dataset.}
            \label{fig:roughness_error_hist_sld}
        \end{figure*}

\begin{figure*}
            \centering
            \includegraphics[]{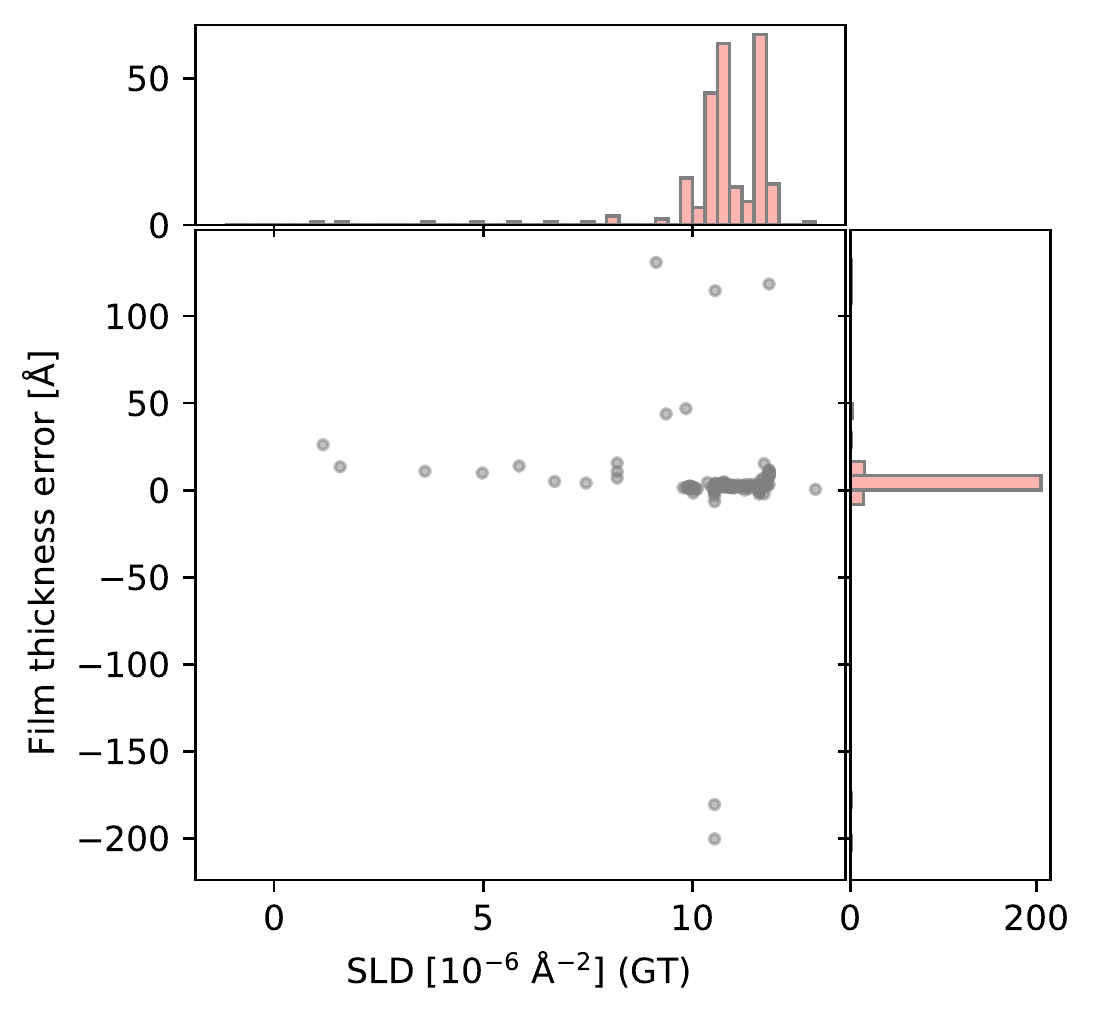}
            \caption{Distribution of the absolute thickness error from the full pipeline fit with respect to the ground truth (GT) SLD. Each dot represents a single curve in the testing dataset.}
            \label{fig:sld_error_hist_thickness}
        \end{figure*}

\begin{figure*}
            \centering
            \includegraphics[]{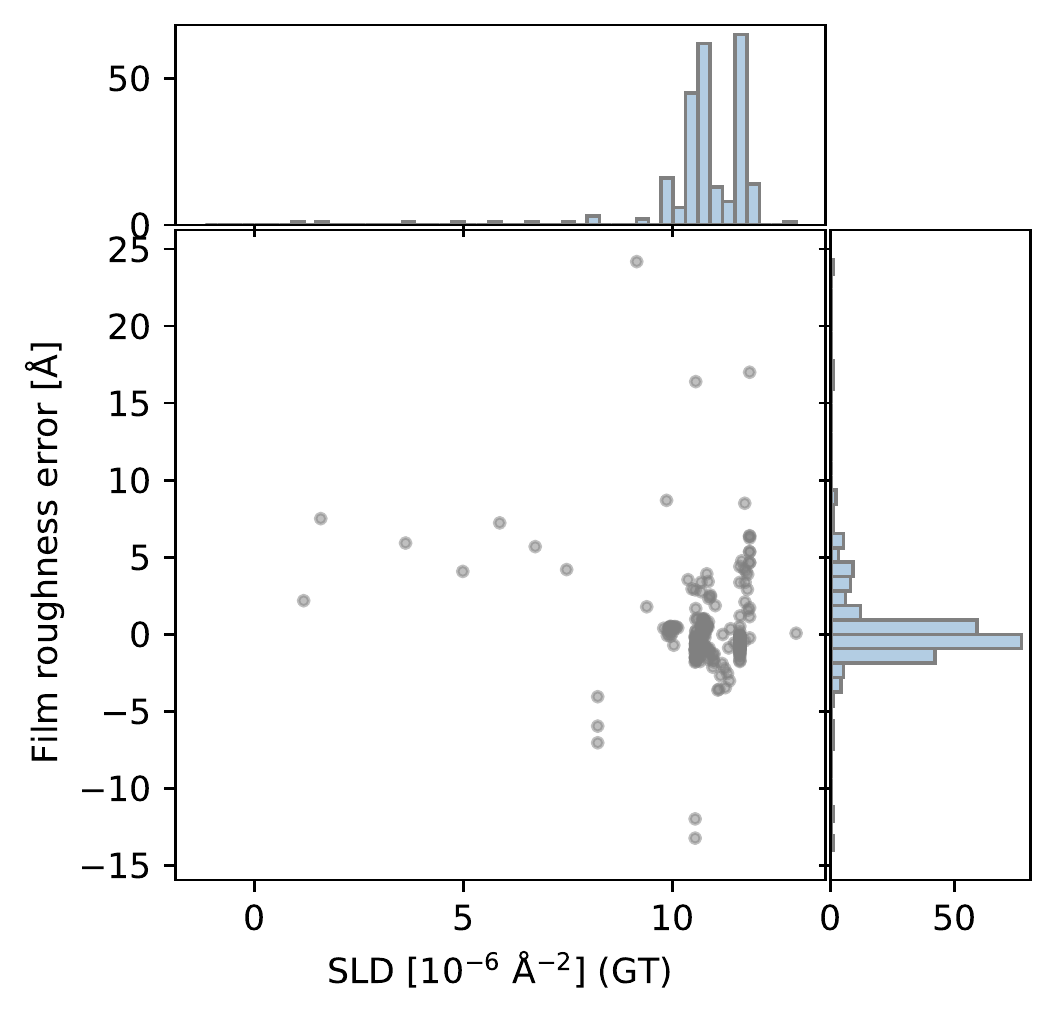}
            \caption{Distribution of the absolute roughness error from the full pipeline fit with respect to the ground truth (GT) SLD. Each dot represents a single curve in the testing dataset.}
            \label{fig:sld_error_hist_roughness}
        \end{figure*}

\begin{figure*}
            \centering
            \includegraphics[]{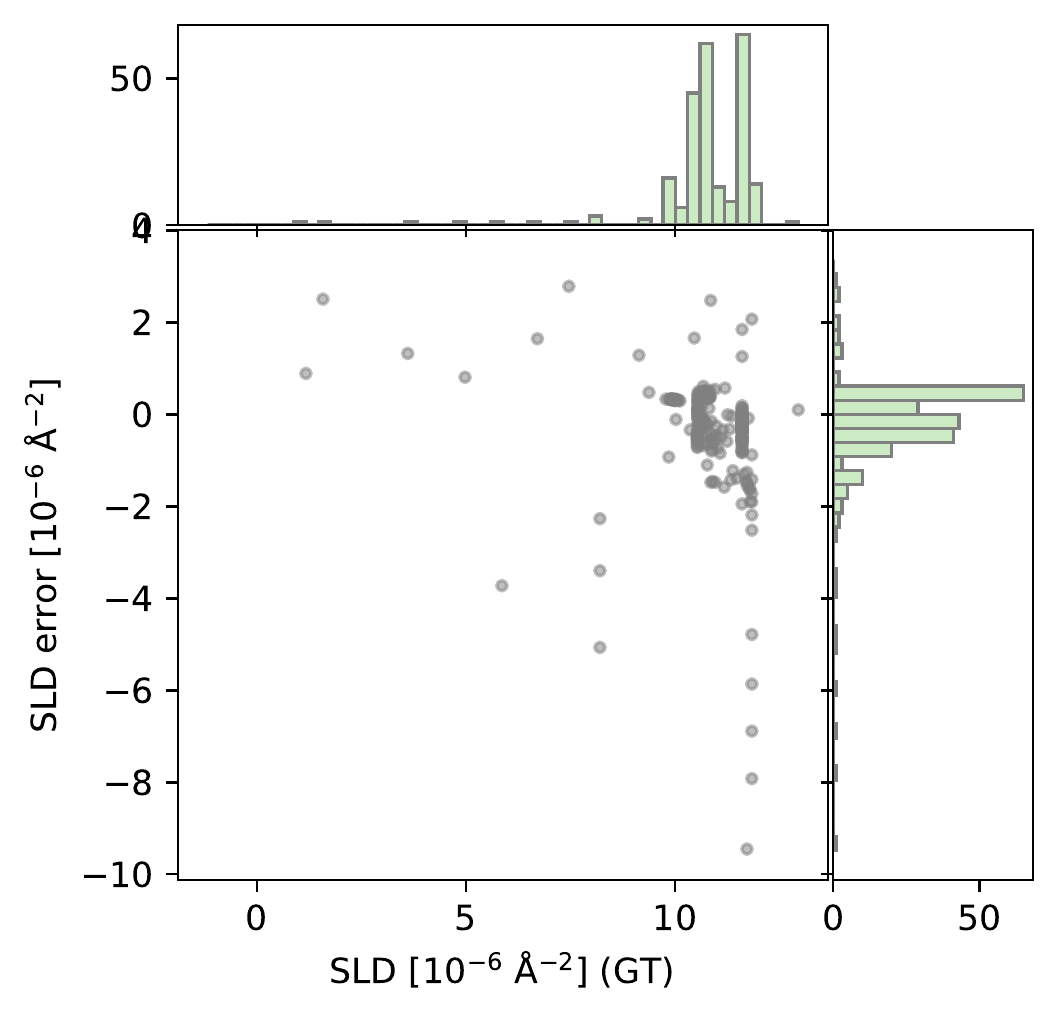}
            \caption{Distribution of the absolute SLD error from the full pipeline fit with respect to the ground truth (GT) SLD. Each dot represents a single curve in the testing dataset.}
            \label{fig:sld_error_hist_sld}
        \end{figure*}

        This section shows detailed histograms of the absolute error distribution of each parameter with respect to the ground truth (GT) of a given parameter, which expands on the condensed form shown in Figure 3 of the main manuscript. Here, only the results of the full pipeline are shown (neural network + q shift + LMS fit). \autoref{fig:thickness_error_hist_thickness}--\ref{fig:thickness_error_hist_sld} show the errors with respect to the GT thickness, \autoref{fig:roughness_error_hist_thickness}--\ref{fig:roughness_error_hist_sld} with respect to the GT roughness and \autoref{fig:sld_error_hist_thickness}--\ref{fig:sld_error_hist_sld} with respect to the GT SLD.
        
        The majority of outliers are due to ambiguous fits (e.g. featureless curves) where multiple parameter combinations lead to a good fit. A common case are very thin films where there are no oscillations visible in the chosen $q$ range.
    
\end{document}